\documentstyle[12pt]{article}
\def\appendixa{
\vskip 1cm
{\bf APPENDIX A:  COUNTER INTEGRAL REPRESENTATION FOR THE
SUM OF THE STRING EIGENFREQUENCIES}
\vskip 1cm
\par
\setcounter{equation}{0}
\def\theequation{A.\arabic{equation}}
}
\def\appendixb{
\vskip 1cm
{\bf APPENDIX B: ASYMPTOTIC EXPANSION OF THE INTEGRAL $I(q)$}
\vskip 1cm
\par
\setcounter{equation}{0}
\def\theequation{B.\arabic{equation}}
}
\renewcommand{\theequation}{\thesection.\arabic{equation}}
\begin{document}
\title{Quark mass correction to the string potential}
\author{G. Lambiase\thanks{Permanent address:
Dipartimento di Fisica Teorica
e S. M. S. A., Universit\'a di Salerno,
84081 Baronissi (SA), Italia},~~~V.V. Nesterenko \\
{\small \it Bogoliubov Laboratory of Theoretical Physics}\\
 {\small \it Joint Institute for Nuclear Research,
Dubna, 141980, Russia}}
\date{}
\maketitle
\begin{abstract}
A consistent method for calculating the interquark potential generated by the
relativistic string with massive ends is proposed. In this approach the
interquark potential in the model of the Nambu--Goto string with point--like
masses at its ends is calculated. At first the calculation is done in the
one--loop
approximation and then the variational estimation is performed. The quark mass
correction results in decreasing the critical distance (deconfinement radius).
When quark mass decreases the critical distance also decreases. For
obtaining a finite result under summation over eigenfrequencies of the
Nambu--Goto string with massive ends a suitable mode--by--mode subtraction
is proposed. This renormalization procedure proves to be completely unique. In
the
framework of the developed approach the one--loop interquark potential in the
model
of the relativistic string with rigidity is also calculated.
\end{abstract}
\vskip0.2cm
PACS number(s): 11.17.+y, 12.38.Aw, 12.40.Aa

\section{Introduction}

Investigation of the quark interaction at large distances is
outside the QCD perturbation theory. Usually, in this field the
lattice simulations and string models are used.

Calculation of the quark interaction in the framework of string
models has rather long history (see, for example,
papers~\cite{OA} -- \cite{POLC} and references therein). In all
these investigations without exception, only the static
interquark potential has been considered. It implies that the
quarks are assumed to be infinitely heavy. Obviously, this
potential, by definition, does not depend on the quark mass.
Assumption about infinitely heavy quarks is rather crude at least
for $u$ and $d$ quarks with masses about 140~MeV that is
significantly less than the characteristic hadronic mass scale
$\sim$ 1~GeV. It is clear, that in a general case the interquark
potential should depend on quark masses. Both the general
approach to this problem in the framework of QCD~\cite{BMPR} and
the numerical calculations of the light and heavy meson spectra
in potential models~\cite{KWON}--~\cite{BUCH} testify to this.
Certainly, in this case one should say not about the static
potential generated, for example, by a relativistic string
connecting quarks but simply about the interaction potential
between quarks having finite mass rather than infinite one.

The aim of the present paper is an attempt to extend the standard
approach to calculation of the interquark potential in the
framework of the string models~\cite{OA},
\cite{KLE}~--~\cite{POLC} to the case of the finite quark mass.
It turns out that this program can be realized. To this end, the
boundary conditions in the string model in question should be
modified and a new renormalization procedure should be developed.

In a proposed approach, a correction to the string potential due
to the finite quark masses is calculated both in the Nambu--Goto
string model and in the Polyakov--Kleinert rigid string model.

In the Nambu--Goto string with massive ends the quark potential
is calculated first in one--loop approximation of perturbation
theory for arbitrary dimension of the space--time $D$ and then
via variational estimation in the limit $(D\,-\,2)\to\infty$. As
knows, the static quark potential generated by Nambu--Goto string
in one--loop approximation is compiled by two terms, linearly
rising confinement potential and universal L\"{u}scher
1/$R$--term~\cite{LUSC}. When the ends of the Nambu--Goto string
are loaded by point--like masses (quarks) then the potential, in
addition to terms mentioned above, acquires correction dependent
on quark mass $m$ and the distance between quarks, $R$. In fact,
$R$ and $m$ are involved in potential correction in the form of a
dimensionless parameter $q\,=\,(M_0/m) M_0\,R$, where $M_0^2$ is
the string tension. As a result, in the correction obtained the
limit of small quark mass $(M_0/m\to\infty)$ is equivalent to the
large $R$ limit $(M_0\,R\to\infty)$, and the limit of heavy
quarks $(M_0/m\to 0)$ is the same as the small $R$ limit. At
large $R$ (or at small $m$) the correction to the quark potential
reduces to the constant determined by $m$ and $M_0^2$ plus terms
of order as $R^{-2}$ and higher. Thus, in the framework of the
perturbative calculation, the universal L\"{u}scher $1/R$--term
is preserved in the Nambu--Goto string with massive ends too. At
small $R$ (or in the case of heavy quarks at string ends) the
mass quark correction to the potential, as one would expect,
vanishes. In our calculation, this is a direct consequence of the
renormalization procedure. It is important that the subtraction
procedure used for this aim proves to be unique.

Variational calculation of the potential generated by the
Nambu--Goto string with massive ends in the limit
$(D\,-2\,)\to\infty$ results in the radical expression (see
Eq.~(4.10)). Quark mass contribution to it can be interpreted as
a substitution of the critical distance $R_c^2\,=\,\pi(D\,-\,2)/
12\,M_0^2$ from the Nambu--Goto string with fixed ends by
quantity $\bar R^2 (m,R)$ dependent on quark mass $m$ and
distance $R$. As a results the radical formula (4.10) has sense
not only at $R\,>\,R_c$, as in the case of the Nambu--Goto string
with fixed ends, but also at $R_c^{eff}\,<R\,<R_c$ where
$R_c^{eff}$ is an effective critical distance determined by
condition $V(R_c^{eff})\,=\,0$. Thus, $R_c^{eff}$ turns out to be
dependent on quark mass $m$ and when $m\to 0$, $R_c^{eff}(m)$
decreases.

In the rigid string model with massive ends the interquark
potential is calculated in one--loop approximation. When
confining to the quadratic approximation in the string action in
this model, the dynamical variables (string position vector) can
be presented as a sum of two terms,
${\bf u}(t,r)\,=\,
{\bf u}_1(t,r)\,+\,{\bf u}_2(t,r)$, where ${\bf u}_1(t,r)$ is a
solution to the Nambu--Goto string with massive ends and ${\bf
u}_2(t,r)$ is additional variable caused by the curvature in the
Polyakov--Kleinert action. It is remarkable that quark masses
only affect on ${\bf u}_1(t,r)$. This essentially simplifies the
problem under consideration and enables us to use directly the
results for potential derived in the Nambu--Goto string with
massive ends. In one--loop approximation, the variables ${\bf
u}_1(t,r)$ and ${\bf u}_2(t,r)$ give additive contribution to the
interquark potential generated by rigid string. It is true both
in the case of the fixed string ends and for the rigid string
with massive ends. As a result, the mass quark correction to the
one--loop potential generated by rigid string is reduced to the
modification of the contribution from the variable ${\bf
u}_1(t,r)$: the one--loop potential in the Nambu--Goto string
with massive ends calculated before should be used here.

In all these calculations we are dealing with an infinite sum of
all eigenfrequencies of the Nambu--Goto string with massive ends.
It substitutes the well know sum over all integer frequencies in
the Nambu--Goto string with fixed or free ends:
$(1/2)\,\sum_{n=1}^{\infty}\,n\,=\,-1/24$. For obtaining a finite
value of this new sum, a mode--by--mode subtraction procedure is
proposed: each initial eigenfrequency is subtracted by the
corresponding frequency of the same string with massive ends
taking the limit $R\to 0$. This prescription turns out to be
completely unique. By making use of the argument principle
theorem, we present this regularized sum in a form of the
integral. Numerical calculation of the subtracted sum over string
eigenfrequencies directly and using for this purpose the integral
representation give the same result.

The layout of the paper is as follows. In Section 2 the quadratic
approximation for the Nambu--Goto string model with massive ends
is developed. Upon linearization of the equations of motion and
boundary conditions, the general solution to them is obtained.
The eigenfrequencies of the string oscillations are determined by
a transcendental equation. Then the canonical quantization is
shortly outlined. In Section 3, the interquark potential
generated by the Nambu--Goto string with massive ends is
calculated in one--loop approximation of the perturbation theory.
In order to remove the divergence, a new subtraction procedure
was proposed. In Section 4, interquark potential, generated by
the Nambu--Goto string with massive ends, is calculated by making
use of a variational estimation of the functional integral in the
limit when $(D\,-\,2)\to \infty$. In Section 5, the rigid string
model with massive ends is treated. By making use of a quadratic
approximation for the Polyakov--Kleinert action, the linear
equations of motion and boundary conditions are derived. Then
canonical quantization of this model is developed. And finally,
interquark potential generated in this string model is calculated
in one--loop approximation. In Conclusion (Section 6) the
obtained results are shortly discussed and possible extension of
them are proposed. Some mathematical details of calculation are
presented in Appendices A and B.

\section{Nambu--Goto string with massive ends}

The action of the Nambu-Goto string with point-like masses
attached to its ends is written in the following way~\cite{BARB}
\begin{equation}
 S\,=\,-\,M_{0}^{2}    \mathop{\int
 \!\!\!\int}\limits_{\Sigma\hspace*{0.25cm}} d\Sigma\,-\,
\sum_{a=1}^{2}m_a
\int\limits_{C_a}^{}ds_a\,{,}
\end{equation}
where $d\Sigma$ is infinitesimal area of the string world
surface, $C_a\;\;(a\,=\,1,2)$ are the world trajectories of the
string massive ends and $M_{0}^2$ is the string tension with the
dimension of the mass squared $(\hbar\,=\,c\,=\,1)$.

For our calculation it will be convenient to use the Gauss
parametrization of the string world surface:

$$x^{\mu}(\xi)=(t,r;x^1(t,r),...,x^{D-2}(t,r))=$$
\begin{equation}
=(\xi^i;{\bf u}(\xi^i)),\quad i=0,1.
\end{equation}
The vector field $u^j(t,r),\,\,j=1,...,D-2$ corresponds to $D-2$
transverse components of $x^{\mu}$, while the $\mbox{\boldmath
$\xi$}=(t,r)$ are the coordinates on the string world sheet. The
infinitesimal area $d\Sigma$ is given by
$d\Sigma\,=\,dt\,dr\,\sqrt{g}$, where $g$ is the determinant of
the induced metric on the world surface of the string,
$g_{ij}\,=\,\partial_{i}x^{\mu}\partial_{j}x_{\mu}$. The metric
of the $D$-dimensional space-time has the signature
$(+,-,\ldots,-)$.

In this parametrization, the induced metric $g_{ij}$ in the quadratic
approximation has the following components
\begin{equation}
g_{ij}\,=\, \delta_{ij}-{\bf u}_{i}{\bf u}_{j}\,{,}
\end{equation}
where ${\bf u} {\bf u}\,= \,\sum_{j=1}^{D-2}u^{j}u^{j}$. From Eq. (2.3)
we obtain
\begin{equation}
g^{ij}\,=\, g^{-1}[(1-{\bf u}_{k}^{2})\,\delta_{ij}\,+\,
{\bf u}_{i}{\bf u}_{j}]\,{,}
\end{equation}
\begin{equation}
g\,=\,\det(g_{ij})\simeq 1\,-\,{\bf u}_{i}^{2}\,{,}
\end{equation}
(here ${\bf u}_0\,=\,\partial{\bf u}/\partial t \,=\,
\dot{\bf u},\quad {\bf u}_1\,=\,
\partial{\bf u}/\partial r\,=\,{\bf u} '$). The
line elements $ds_{a},\; a=1,2,$ take the form:
\begin{equation}
ds_a\simeq [1-\frac{1}{2}{\dot{\bf u}}^2(t,\, r_a)]\,dt\,{.}
\end{equation}
After neglecting unimportant constants, the action (2.1) becomes
\begin{eqnarray}
S&\simeq & -\,M_0^2\,(t_2-t_1)R\,+
\, \frac{M_{0}^{2}}{2}\int\limits_{t_1}^{t_2}\!\!dt\!\!
\int\limits_{0}^{R}\!\!dr\left [
\dot{\bf u}^2(t,r)-{\bf u}'^2(t,r)\right ]\,+ \nonumber \\
&&+\sum_{a=1}^{2}\frac{m_a}{2}\int\limits_{t_1}^{t_2}dt\,
{\dot{\bf u}}^2(t,r_a),\quad
r_1\,=\,0,\quad r_2\,=\,R\,{.}
\end{eqnarray}
The last two terms in Eq.~(2.7), important for deriving
dynamical equations, can
be rewritten as follows
\begin{equation}
\bar S \simeq \frac{M_{0}^2}{2}
\int\limits_{t_1}^{t_2}dt\int\limits_{0}^{R}dr\left [
\dot{\bf u}^2(t,r)\varepsilon(r)-{\bf u}^{'2}(t,r)\right ]\,{,}
\end{equation}
where $\varepsilon(r)$ is the mass density distribution along the string
\begin{equation}
\varepsilon (r)\,=\,1\,+\,\frac{m}{M_0^2}\left [\delta(r)\,+\,
\delta(R\,-\,r)
\right]\,{.}
\end{equation}
For semplicity we assume that the string ends are loaded with
equal masses, $m_1\,=\,m_2
\,=\,m$.

Equations of motion and boundary conditions can be deduced from (2.7) or
(2.8) by relations
\begin{equation}
\frac{\partial}{\partial t}\left (\frac{\partial L_{str}}
{\partial\dot{\bf u}}\right)+
\frac{\partial}{\partial r}\left (\frac{\partial L_{str}}{\partial
{\bf u}'}\right)\,=\,0\,{,}
\end{equation}
and
\begin{equation}
\frac{\partial L_{str}}{\partial{\bf u}'}-(-1)^{a}
\frac{\partial}{\partial t}
\frac{\partial L_{a}}{\partial\dot{\bf u}_{a}}\,=\,0\,{,}
\quad a=1,2\,{,} \quad r_a=0,R.
\end{equation}
$L_{str}$ is the string Lagrangian density and $L_a,\, a=1,\,2$
are the Lagrangians of the massive ends of the string in Eq. (2.7).
Equations of motion are given by
\begin{equation}
\Box{\bf u}=0\,{,}
\end{equation}
where $\Box=\partial^2/\partial{t^2}-\partial^2/\partial{r^2}$,
while for boundary conditions we find~\cite{VVN}
\begin{equation}
m\ddot{\bf u}\,=\,M_{0}^{2}{\bf u}'\,{,}\qquad r\,=\,0\,{,}
\end{equation}
\begin{equation}
m\ddot{\bf u}\,=\,-M_{0}^{2}{\bf u}'\,{,}\,\qquad r\,=\,R\,{.}
\end{equation}
Equation (2.12) admits the following solutions:
\begin{equation}
u^j (t,r) \sim \alpha^j  \exp[i(\omega/R)t]\, u(r)\,{,}\qquad
j\,=\,1,2,\ldots, D-2\,{,}
\end{equation}
in which the string length $R$ has been introduced in order that
$\omega$ be dimensionless. By substituting (2.15) into (2.12),
(2.13) and (2.14) one obtains
\begin{equation}
{\bf u}''(r)+\frac{\omega^2}{R^2}\,{\bf u}(r)=0\,{,}
\end{equation}
\begin{equation}
\omega^2 \,{\bf u}(0)=-qR\,{\bf u}'(0)\,{,}
\end{equation}
\begin{equation}
\omega^2 \,{\bf u}(R)=qR\,{\bf u}'(R)\,{,}
\end{equation}
where $q$ is a dimensionless parameter
\begin{equation}
q=\frac{M_{0}^{2}R}{m}\,{.}
\end{equation}
The differential equations (2.16)--(2.18) are linear, therefore
each component of the transverse oscillation of the string
satisfies the same equations. Hence, we can write the general
solution as a superposition of plane wave solutions
\begin{equation}
u^{j}(t,r)=i\,\frac{\sqrt{2}}{M_0}\sum_{n\not= 0}^{}
\exp[-i(\omega_{n}/R) t]
\,\frac{\alpha^{j}_{n}}{\omega_{n}}\,u_{n}(r)\,{,}
\quad j=1,...,D-2\,{,}
\end{equation}
where the amplitudes $\alpha^{j}_{n}$ (Fourier coefficients) obey
the usual rule of complex conjugation,
$\alpha_{n}=\alpha_{-n}^{*}$, in order that $u_{n}^{j}$ be real.
The eigenfunctions $u_{n}(r)$ in (2.20) are defined by
\begin{equation}
u_n (r)\,=\,N_{n}\,\left[\cos\left({\omega_n\frac{r}{R}}\right)\,-\,
\frac{\omega_n}{q}\sin\left({\omega_n\frac{r}{R}}\right)\right] \,{,}
\end{equation}
where $N_n$ are normalization constants and the
eigenfrequencies $\omega_n$
are the roots of the transcendental equation
\begin{equation}
\tan{\omega}\,=\,\frac{2q\omega}{\omega^{2}-q^2}\,{.}
\end{equation}

On the $\omega$-axis these roots are placed symmetrically around
zero. Hence they can be numbered in the following way
$\omega_0\,=\,0,\,\omega_{-n}\,=\,-\omega_n,\,n=1,2,\ldots$.
Therefore it will be sufficient to consider only the positive
roots. The eigenfunction $u_n(r)$ obey the orthogonality
conditions with the weight function $\varepsilon(r)$~\cite{MFH}
\begin{equation}
\int\limits_{0}^{R}dr u_{n}(r)\,u_{m}(r)\varepsilon(r)\,=\,
R\,\delta_{nm}\,{,}
\end{equation}
while the functions $u_{n}'(r)$ satisfy the usual orthogonality
conditions
\begin{equation}
\int\limits_{0}^{R}dr u^{'}_{n}(r)\,u^{'}_{m}(r)\,=\,
\frac{\omega_n^2}{R}
\,\delta_{nm}\,{,}
\end{equation}
where the eigenfrequencies $\omega_n$ are solutions of Eq. (2.22).

The density of the canonical momentum $p^{j}(t,r)$ is defined in
     standard way
\begin{equation}
p^{j}(t,r)\,=\,\frac{\partial L_{tot}}{\partial
{\dot{u}^j}}=M_{0}^{2}\dot{u}^{j}
(t,r)\,\varepsilon(r)\,{,}
\end{equation}
in which $L_{tot}$ is the Lagrangian density in action (2.8). From
(2.25) the total momentum of the string is given by
\begin{equation}
P^{j}(t)\,=\,\int\limits_{0}^{R}dr\, p^{j}(t,r)\,{.}
\end{equation}
Obviously, in the problem under consideration, we can put $P_j\,=\,0$.
The canonical Hamiltonian is defined by
$$H\,=\,\int\limits_{0}^{R}\!\!dr\,[{\bf p}(t,r)\,\dot{\bf u} (t,r)\,-
\,L_{tot}]\,=$$
\begin{equation}
=\,\frac{M_{0}^{2}}{2}\int\limits_{0}^{R}dr\,
[\dot{\bf u}^{2}(t,r)\varepsilon(r)+
{\bf u}^{'2}(t,r)]\,{,}
\end{equation}
In terms of the amplitudes $\alpha_{nj}$ it reads
\begin{equation}
H\,=\,\frac{1}{R}\,\sum_{n=1}^{\infty}\sum_{j=1}^{D-2}\left(\alpha_{n}^j\,
\alpha_{n}^{j+}
+\alpha_{n}^{j+}\,\alpha_{n}^j\right)\,{.}
\end{equation}

When we quantize, $u^{j}(t,r)$ and its conjugate momentum
$p^{j}(t,r)$ become operators with canonical commutation
relations
\begin{equation}
[u^i (t,r),\,p^j (t,r')]\,=\,i\,\delta^{ij}\,\delta(r-r')\,{,}
\end{equation}
This implies that the Fourier coefficients become
operators and satisfy the relations
\begin{equation}
[\alpha^{i}_{n},\alpha_{m}^{j}]\,=\,\omega_{n}\delta^{ij}\,
\delta_{n+m,0}\,{,}
\end{equation}
$$ i,j\,=\,1,\ldots,D-2\,{,}\quad n,m\,=\,\pm 1,\pm 2,\ldots\,{.}$$

The creation and annihilation operators, respectively
$a_{n}^{+}$ and $a_{n}$,
in Fock space are introduced in usual way
\begin{equation}
\alpha_{n}^{j}\,=\,\sqrt{\omega_{n}}\,a_{n}^{j}\,{,}\qquad
\alpha_{n}^{j\,+}\,=\,\sqrt{\omega_{n}}\,a_{n}^{j\,+}\,{,}
\end{equation}
\begin{equation}
[a_{n}^{i},a_{m}^{j\,+}]\,=\,\delta^{ij}\,\delta_{nm}\,{,}
\quad n,m\,=\,1,2,\ldots\,{,}
\end{equation}
and through them, the Hamiltonian (2.28) takes the form
\begin{equation}
H\,=\,\frac{1}{R}\,\sum_{n=1}^{\infty}\sum_{j=1}^{D-2}
\,\omega_n\,a_{n}^{j+}\,a_{n}^{j}+
\frac{D-2}{2}\,\frac{1}{R}\,\sum_{n=1}^{\infty}\,\omega_{n}\,{.}
\end{equation}
The last term in (2.33) is the usual Casimir energy~\cite{MT,PMG}.

\section{One--loop potential generated by Nambu--Goto string with
massive ends}
\setcounter{equation}{0}

   In this section we shall investigate of the interquark
potential arising from the action (2.1), via the perturbation
calculation. For this purpose the Euclidean version of the model
under consideration should be treated.

Up to the second order in ${\bf u}$ the string action (2.1) in
Euclidean space is given by $(t_{1}=0,\, t_{2}=T)$
\begin{equation}
S_{E}\,=\,M_{0}^{2}\,\int\limits_{0}^{T}\!\! dt\int
\limits_{0}^{R}\!\!dr
\left[1\,+\,\frac{1}{2}{\bf u}\,\partial^{2}{\bf u}\right]\,+\,
\sum_{a=1}^{2}\,
\frac{m_a}{2}\,\int\limits_{t_1}^{t_2}\!\!dt\,{\dot{\bf u}}^2(t,r_a)\,{,}
\end{equation}
$$ r_1\,=\,0,\qquad r_2\,=\,R\,{,}$$
where
$\partial^{2}\,=\,\partial^{2}/\partial{t^2}+\partial^{2}/\partial{r^2}$.
It generates one--loop Feynman diagrams in perturbation theory.
The potential $V(R)$ between massive quark separated by a
distance $R$ is defined in terms of functional integral as
follows~\cite{KW,LSW}:
\begin{equation}
e^{-T\,V(R)}\,=\,\int\limits_{}^{}[D{\bf u}]\,e^{-S_{E}[{\bf u}]}
\,{,}\qquad
T\rightarrow\infty\,{.}
\end{equation}
 From action (3.1), after functional integration, the interquark potential
reads
\begin{equation}
V(R)=\lim_{T\rightarrow\infty}\,\frac{1}{T}\,\left[\int
\limits_{0}^{T}\!\!
dt\int\limits_{0}^{R}\!\!dr\,M_{0}^{2}\,+\,
\frac{D-2}{2}\,\mbox{Tr}
\ln G^{-1}\right]\,{.}
\end{equation}

In Eq.~(3.3) $G^{-1}\,=\,\partial^2\,$ is the operator generated
by the quadratic part of the action $S_{E}$ and $G$ is the
Euclidean Green function
\begin{equation}
\partial^2\,G(x,x')\,=\,\delta(x,x')\,{.}
\end{equation}

In the momentum space this function is written as
\begin{equation}
G(k_{0},k_{n})\,=\,\frac{1}{k_{0}^{2}+k_{n}^{2}}\,{,}
\end{equation}
where $k_{n}$ are admissible values of the wave vector for the field
${\bf u}(t,r)$,
determined by the boundary conditions (2.13) and (2.14)
\begin{equation}
k_{n}\,=\,\frac{\omega_n}{R}\,{.}
\end{equation}
As before $\omega_n$ are the roots of frequency equation (2.22).
By using the definition of the Tr in the momentum space
\begin{equation}
\mbox{Tr}\ldots\,=\,T\!\!\int\limits_{-\infty}^{+\infty}
\!\!\frac{d\omega}{2\pi}
\sum_{n}^{}\ldots\,{,} \qquad \omega=k_{0} \,{,}
\end{equation}
where the sum is spread over all the discrete values of
the component $k_{n}$,
we can present the functional trace in Eq. (3.3) in the following way
\begin{equation}
\mbox{Tr}\,\ln G^{-1}\,=\,T\,\sum_{n}^{}
\int\limits_{-\infty}^{+\infty}\!\!
\frac{d\omega}{2\pi}\,\ln(\,\omega^2\,+\,k_{n}^{2})\,{.}
\end{equation}
In the analytical regularization the $\omega$-integration can be
performed by the formula~\cite{KCN}
\begin{equation}
\int\limits_{-\infty}^{+\infty}\!\!\frac{d\omega}{2\pi}\,
\ln(\omega^2\,+
\,a^2)\,=\,\sqrt{a^2}\,{.}
\end{equation}

    It follows that up to the one-loop level the interquark
potential (3.3) is given by
\begin{equation}
V(R)\,=\,M_{0}^{2}\,R\,+\,\frac{D-2}{R}\,v(q) \,{,}
\end{equation}
in which
\begin{equation}
v(q)\,=\,\frac{1}{2}\,\sum_{n=1}^{\infty}\,\omega_{n}(q)\,{,}
\end{equation}
and $\omega_n(q)$ are the roots of Eq. (2.22).

The sum in (3.11) diverges~\cite{MT,PMG} so that it is necessary
to develop a subtraction procedure physically acceptable in order
to get the renormalized function $v^{ren}(q)$ in a unique way. We
propose here a mode-by-mode subtraction procedure. To this end we
consider the asymptotic of eigenfrequencies $\omega_n(q)$ for
large $n$ but fixed $q$~\cite{FWJ}
\begin{equation}
\omega_{n}(q)\,\sim \,n\,\pi\,+\,\frac{2q}{n\pi}\,-\,
\left(1\,+\,\frac{q}{6}\right)\,\frac{(2q)^2}{(n\pi)^3}+O(n^{-5})
\,{,}\quad n \to\infty\,{.}
\end{equation}
The divergence of the sum (3.11) is due to the first two terms in (3.12).
At the same time we consider the asymptotic forms of Eq. (2.22) at $q \to 0$
\begin{equation}
\tan \bar\omega \,=\,\frac{2q}{\bar\omega}\,{,}
\end{equation}
and at $q \to \infty$
\begin{equation}
\cot \tilde\omega\,=\,-\frac{q}{2\tilde\omega}\,{.}
\end{equation}
The asymptotic of the roots of these equations for large $n$
and fixed $q$
are given, respectively, by
\begin{equation}
\bar\omega(q)\,\sim\,n\,\pi\,+\,\frac{2q}{n\pi}\,-\,\left(1\,
+\,\frac{2q}{3}
\right)\,\frac{(2q)^2}{(n\pi)^3}\,
+\,O(n^{-5})\,{,}\qquad n \to \infty\,{,}
\end{equation}
and
\begin{equation}
\tilde\omega(q)\,\sim\,n\,\pi\,+\,\frac{q}{2n\pi}\,+\,
\frac{q^3}{24\,n^3\pi^3}\,+\, O(n^{-5})\,{,} \quad n\to \infty\,{.}
\end{equation}
Thus we see that in order to remove the divergencies in sum
(3.11) we must substitute $\omega _n$ by $\omega_n \,-\,\bar
\omega_n$. The roots of Eq.~(3.14) cannot be used for this
purpose in view of their asymptotic (3.16). Therefore the
subtraction procedure proposed here is unique.

In terms of the function $v(q)$ the proposed renormalization
procedure is defined in the following way
\begin{equation}
v^{ren}(q)\,=\,v(q)\,-\overline{\lim}_{q\to 0}\,v(q)\,+\,v_0\,{.}
\end{equation}
The symbol $\overline{\lim_{q\to 0}}\,v(q)$ means that we must
take the asymptotic expression of the function $v(q)$. Analogous
renormalization scheme is used by calculating the Casimir energy
in some field models with special boundary conditions. We have
introduced here a constant $v_0$ in order to satisfy the
requirement that at $m\,=\,\infty$ Eq.~(3.17) must give
$v^{ren}(q)$ for string with fixed ends, i.e.
\begin{equation}
v^{ren}(q)\,=\,v(q)\,-\,\overline{\lim_{q\to 0}}\,v(q)\,
+\,v(q)\vert_{m\,=\,
\infty}\,{.}
\end{equation}
The frequencies of the string with fixed ends are
\begin{equation}
\omega_n\,=\,n\,\pi\,{.}
\end{equation}
Therefore the last terms in (3.18) is defined by
\begin{equation}
v_0\,=\,v(q)\vert_{m\,=\,\infty}\,=\,\frac{\pi}{2}\,
\sum_{n=1}^{\infty}\,n\,{.}
\end{equation}
This sum can be renormalized by making
use of the Riemann zeta-function $\zeta (s)\,=\,
\sum_{n=1}^{\infty}\,n^{-s}$:
\begin{equation}
v(q)\vert_{m\,=\,\infty}\,=\,\frac{\pi}{2}\,
\sum_{n=1}^{\infty}\,n\,=\,
\frac{\pi}{2}\,\zeta(-1)\,=\,-\frac{\pi}{24}\,{.}
\end{equation}
The final expression for the renormalized function $v^{ren}(q)$
     is given by
\begin{equation}
v^{ren}(q)\,=\,-\,\frac{\pi}{24}\,+\,\frac{1}{2}\,\sum_{n=1}^{\infty}\,
\left[\,\omega_{n}(q)\,-\,\bar\omega_n(q)\,\right]\,{.}
\end{equation}

This function has been calculated numerically.
As shown in Fig. 1, $v^{ren}(q)$ is a monotone increasing function
of the dimensionless parameter $q$.

The interquark potential (3.10) in terms of the function
     $v^{ren}(q)$ reads
\begin{equation}
V(R)\,=\,M_{0}^{2}\,R\,+\,\frac{D-2}{R}\,v^{ren}(q)\,{.}
\end{equation}
Equation (3.23) shows that the potential generated by string with
massive ends depends on the quarks masses through the
dimensionless parameter $q\,=\,M_{0}^{2}R/m$. Taking into account
this fact we can give here a very clear physical meaning of the
subtraction procedure proposed above for obtaining renormalized
quantum correction to the potential (3.22). The limit $q\to 0$
may be obtained by putting $R\to 0$. Therefore, $v^{ren}(q)$ is
determined as a difference of $v(q)$ in two points: $r\,=\,R$ and
$r\to 0$. In the same way, the potential (3.23) is equal to the
work that should be done for removing quarks connected by string
to the distance $R$, provided that at the beginning they were
very closed to each other.

Certainly, this renormalization procedure should be followed by
substitution for $M_0^2$, the renormalized (physical) value of
the string tension. For simplicity we do not introduce new
notations and will keep this point in mind.

Now we propose a different way for calculating the sum in Eq.
(3.22). It utilize an analytical method that employes the
following integral formula from the complex analysis~\cite{LS}.

Let us consider an analytical function $f(z)$ with zeroes of
order $n_{k}$ at points $z\,=\,a_{k}$ and with poles of order
$p_l$ at points $z\,=\,b_l$ in a region bounded by a contour $C$.
 From Cauchy's theorem it follows that
\begin{equation}
\frac{1}{2\pi i}\,\oint\limits_{C}^{}\!dz\,z\,\frac{f'(z)}{f(z)}\,=\,
\frac{1}{2\pi i}\,\oint\limits_{C}^{}\!dz\,z\,[\ln\,f(z)]'\,=\,
\sum_{k}^{}\,n_k a_k\,-\,\sum_{l}^{}\,p_l b_l\,{.}
\end{equation}

In order to obtain the renormalized function $v^{ren}(q)$ through
Eq.~(3.24), one has to choose an appropriate expression for the
function $f(z)$. Equation (3.18) implies that $f(z)$ should be
taken in the following form
\begin{equation}
f(\omega)\,=\,\frac{\omega^2\,-\,q^2\,-\,2\,q\,\omega\,
\cot\omega}{\omega^2\,
-\,2\,q\,\omega\,\cot\omega}\,\sin\omega \, {.}
\end{equation}
Here the numerator is in fact Eq. (2.22) and the denominator is the
same equation in the limit $q\rightarrow 0$ (Eq. (3.13)).
The multiplier $\sin\omega$ is introduced to take into account
the frequencies (3.19).

When applying the formula (3.24) to nominator, denominator and
multiplier $\sin\omega$ in (3.25) separately with the counter $C$
encircling the real positive semiaxis in the complex plane
$\omega$, the contribution of the poles into (3.24) is obviously
absent. Now we deform the contour $C$ so that it transforms into
the imaginary axis on the complex plane. After integrating by
parts, the final expression for the renormalized function
$v^{ren}(q)$ becomes (Appendix A)
\begin{equation}
v^{ren}(q)\,=\,-\frac{\pi}{24}\,+\,I(q)\,{,}
\end{equation}
where
\begin{equation}
I(q)\,=\frac{1}{2\pi}\,\int\limits_{0}^{\infty}
\!\!dy\ln\,\left[1+\frac{1}{(y/q)^2\,+\,2(y/q)\coth\,y}\right]\,{.}
\end{equation}
The integral $I(q)$ has been calculated numerically for different
values of the parameter $q$. The results obtained for
$v^{ren}(q)$ by making use of (3.22) and (3.26), (3.27) are
certainly the same, (see Fig. 1).

Finally we analyze the asymptotic expansion of the potential
$V(R)$ in the limit $q\to\infty$ or $R\to\infty$. In this region
the renormalized function $v^{ren}(q)$, Eq.~(3.26), is given by
(see Appendix B)
\begin{equation}
v^{ren}(q)\,\simeq \,-\frac{\pi}{24}\,+\,\frac{\ln 2}{\pi}\,q\,
+\,\frac{1}{2\pi q}
\,{,}\quad q\,=\,\frac{M_0^2\,R}{m}\,{,} \quad q\to\infty \,{.}
\end{equation}
At large $R$ the interquark potential (3.23) can be written up to
second order in $1/R$ as follows
$$ V(R)\,\simeq \,M_{0}^2\,R\,+\,\frac{(D-2)\,\ln 2}{\pi}\,
\frac{M_{0}^2}{m}\,-\,
\frac{\pi(D-2)}{24}\,\frac{1}{R}\,+$$
\begin{equation}
+\,\frac{D-2}{2\pi}\,\frac{m}{M_{0}^2}\,
\frac{1}{R^2}\,+\,O(R^{-4})\,{,} \quad R\to \infty\,{.}
\end{equation}

The first terms in (3.29) is the usual linearly rising term
$M_0^2\,R$ (confining potential). The second term is a constant
dependent on the quark mass. The third term is the universal
L\"{u}scher $1/R$--term and the last one is a correction to the
string potential due to the finite quark mass.

\section{Variational estimation of the string potential}
\setcounter{equation}{0}

In preceding Section the string potential has been calculated by
using the perturbative theory for an arbitrary dimension of
space-time $D$. Otherwise this potential can be derived in the
limit $D\to\infty$ by making use of the variational estimation of
the functional integral~\cite{OA},~\cite{PIAL}--\cite{POLC}.

Let us turn to the initial equation (3.2) determining the
string potential
\begin{equation}
e^{-T\,V(R)}\,=\,\int[D{\bf u}]\,e^{-S_E [{\bf u}]}
\,{,}\quad T\to\infty\,{,}
\end{equation}
where $S_E$ is the Euclidean action
\begin{equation}
S_E\,=\,M_{0}^{2}\,\int\limits_{0}^{T}\!\!dt\int\limits_{0}^{R}\!\!dr\,
\sqrt{\det (\delta_{ij}\,+\,\partial_i{\bf u}\,
\partial_{j}{\bf u})}\,+\,
\sum_{a=1}^{2}\,m_a\,\int\limits_{0}^{T}\!\!dt\,\sqrt{1\,+\,
\dot{\bf u}^2(t,r_a)}
\,{,}
\end{equation}
$$ r_1\, =\,0,\qquad r_2\,=\,R\,{.}$$
The $1/(D-2)$-expansion is carried out in standard way~\cite{OA}.
Let us introduce the composite field $\sigma_{ij}$ for
$\partial_i{\bf u}\,\partial_j {\bf u}$ and constrain
$\sigma_{ij}\,=\,\partial_i{\bf u}\,\partial_j{\bf u}$ through
the Lagrange multiplier $\alpha^{ij}$. By using the exponential
parametrization of the $\delta$-function, with the understanding
that the $\alpha^{ij}$ functional integrals run from $-i\infty$
to $+i\infty$, and calculating the ${\bf u}$-functional integral,
Eq. (4.1) becomes
\begin{equation}
e^{-T\,V(R)}\,=\,\int\limits_{}^{}[D\alpha]\,[D\sigma]\,
e^{-\,\bar S_{E}[\alpha,\sigma]}\,{,}\quad T\to\infty\,{,}
\end{equation}
where
$$\bar S_{E}\,=\,M_{0}^2\,\int\limits_{0}^{T}\!\!
dt\int\limits_{0}^{R}\!\!dr
\,\left[\sqrt{\det(\delta_{ij}\,+\,\sigma_{ij})}\,-\,
\frac{1}{2}\,\alpha^{ij}\,
\sigma_{ij}\right]\,+$$
\begin{equation}
\,+\,\frac{D-2}{2}\,\mbox{Tr}\ln (-\partial_i\alpha^{ij}\partial_j)
\,{.}
\end{equation}
The boundary terms in (4.2) should be taken into account under
finding the eigenvalues of the differential operator
$-\,\partial_i\,\alpha^{ij}\, \partial_j$ in (4.4). The
$1/(D-2)$-expansion is generated by expanding the action (4.4)
around its stationary point. We shall assume that at the
stationary point $\alpha^{ij}$ and $\sigma_{ij}$ are diagonal
matrices, $\bar\alpha$ and $\bar\sigma$ respectively, independent
on $t$ and $r$~\cite{OA}. From Eqs. (3.6), (3.7) and (3.9) we
obtain
$$
\mbox{Tr}\ln(-\partial_i\alpha^{ij}\partial_{j})\,=
\,T\,\sum_{n}^{}\,\int\limits_{-\infty}^{+\infty}\!
\frac{d\omega}{2\pi}\,\ln(\alpha^{00}\,\omega^2\,+\,
\alpha^{11}\,k_n^2)\,=$$
\begin{equation}
=\,\frac{2T}{R}\,\sqrt{\frac{\alpha^{11}}{\alpha^{00}}}\,v(q)\,{,}
\end{equation}
where $v(q)$ is defined by (3.11). Now, action (4.4) can
be rewritten as
$$
\bar S_E\,=\,M_{0}^2\,R\,T\,\left[\sqrt{1\,+\,
\sigma_0}\sqrt{1+\sigma_1}\,-\,
\frac{1}{2}\,(\alpha^0\sigma_0\,+\,\alpha^1\,
\sigma_1)\right]\,+$$
\begin{equation}
+\,(D-2)\frac{T}{R}\sqrt{\frac{\alpha^1}{\alpha^0}}\,v(q)\,{,}
\end{equation}
with $\alpha^{ii}\,=\,\alpha^i,\,\sigma_{ii}\,=\,\sigma_i,
\,i=0,1$.
Equating to zero the
variations of the action (4.6) with respect to
 $\sigma_0,\,\sigma_1,\,\alpha^0,\,
\alpha^1$ and  solving   the corresponding set of
algebraic equations allow  us to
find the stationary points
\begin{equation}
\bar\alpha^0\,=\,\sqrt{1\,-\,2\lambda}\,{,}
\quad\bar\alpha^1\,=\,\frac{1}
{\sqrt{1\,-\,2\lambda}}\,{,}
\end{equation}
\begin{equation}
\bar\sigma_0\,=\,\frac{\lambda}{1\,-\,2\lambda}\,{,}
\quad\bar\sigma_1\,=\,
-\,\lambda\,{,}
\end{equation}
where  the dimensionless function $\lambda(q)$ is defined by
\begin{equation}
\lambda(q)\,=\,-\,\frac{D-2}{M_0^2\,R^2}\,v^{ren}(q)\,{.}
\end{equation}
Here we have used the renormalized function $v(q)$ from Eq.~(3.22).

Putting (4.7) and (4.8) into (4.6) yields the string potential to
leading order in $1/(D-2)$:
\begin{equation}
V(R)\,=\,M_0^2\,R\,\sqrt{\,1\,-\,\frac{R_c^2}{R^2}\,
\left(1\,-\,\frac{24}{\pi}
\,I(q)\right)}\,{,}
\end{equation}
where $R_c^2\,=\,\pi (D-2)/12M_0^2$ and $I(q)$ is defined in
Eq.~(3.27).
The quark mass contribution in this formula can be interpreted as
a substitution of $R_c^2$ from the Nambu--Goto string with fixed
ends by function $\bar R^2(m,R)$ dependent on quark mass $m$ and
distance $R$:
\begin{equation}
\bar R^2(m,R)\,=\,R_c^2\,\left[1\,-\,
\frac{24}{\pi}\,I\left(\frac{M_0^2\,R}{m}
\right)\right]\,{.}
\end{equation}
This results in an extension of the applicability of Eq.~(4.10)
to $R\,<\,R_c$. More precisely, formula (4.10) has sense not only
for $R\,>\,R_c$, that takes place for the string with fixed ends,
but also in the region
\begin{equation}
R_c^{eff}\,<\,R\,<\,R_c\,{,}
\end{equation}
where $R_c^{eff}$ is defined by equation
\begin{equation}
\bar R^2(m,R_c^{eff})\,=\,(R_c^{eff})^2\,{.}
\end{equation}
At $R\,=\,R_{c}^{eff}$ the interquark potential (4.10) vanishes.
Now critical distance $R_{c}^{eff}$ is determined by the quark
mass $m$ and when $m\to 0$ $R^{eff}(m)$ also decreases. In Fig. 2
the dimensionless potential $V(R)/M_0$ as a function of the
dimensionless distance $\rho\,=\,M_0\,R$ is plotted for different
values of the ratio $\mu\,=\,M_0/m$.

At the end of this Section we present an asymptotic expression
for the interquark potential (4.10) at $R\to \infty$. The
asymptotic expression for function $I(q)$ when $q\to \infty$, (or
$R\to\infty$), is given by (B.5)
\begin{equation}
I(q)\simeq \frac{\ln 2}{\pi}\,\frac{M_0^2\,R}{m}\,+\,\frac{1}{2\pi}\,
\frac{m}{M_0^2\,R}\,{,}\quad R\to\infty\,{.}
\end{equation}
Taking into account (4.14), we obtain
$$
V(R) \sim M_0^2\,R\,+\,\frac{(D\,-\,2)\,\ln 2}{\pi}\,
\frac{M_0^2}{m}\,-\,
\frac{\pi\,(D\,-\,2)}{24\,R}\,
\left[1\,+\,\frac{12\,(D\,-\,2)\,(\ln 2)^2}{\pi^3}\,
\frac{M_0^2}{m^2}\right]
\,+
$$
\begin{equation}
+\,\frac{D-2}{2\pi\,R^2}\frac{m}{M_0^2}\,\left[1\,
+\,\frac{\pi(D-2)\,\ln 2}{12}\,
\frac{M_0^2}{m^2}\right]\,+\,O(R^{-3})\,{,}\qquad R\to \infty
\,{.}
\end{equation}

The interquark potential calculated via variational approach, at
large values of $R$ has some additional terms as compared with
the potential (3.29), obtained by perturbative method. In
particular, it turns out that the L\"{u}scher $1/R$-term
\begin{equation}
\frac{ \pi(D\,-\,2)}{24\,R}
\end{equation}
should be now substituted by
\begin{equation}
\frac{\pi\,(D\,-\,2)}{24\,R}\left[1\,+\,\frac{12\,(D\,
-\,2)\,(\ln 2)^2}{\pi^3}\,
\frac{M_0^2}{m^2}\right]\,{.}
\end{equation}

\section{Quark mass corrections to the rigid string potential}
\setcounter{equation}{0}

In this Section we calculate the mass quark corrections to the
one--loop interquark potential in the framework at the rigid
string model ~\cite{KLE,POL}. As known, this model can be treated
as an effective one taking into account the finite thickness of
gluonic tube ~\cite{RG}--\cite{INSKI}. The basic aim of this
calculation is to show the principal applicability of the
proposed method to the rigid string model with massive ends.
Variational estimation of the interquark potential in the
framework of this model will be published elsewhere.

The Polyakov-Kleinert action for rigid string with massive ends
has the form
\begin{equation}
S\,=\,-M_0^2\,\mathop{\int\!\!\!\int}_{\Sigma\hspace*{0.25cm}}
d\Sigma\,
\sqrt{-g}\left[\,1\,-\,\frac{\alpha}{2}\,r_s^2\,
\Delta x^{\mu}\,\Delta x_{\mu}
\right]\,-\,\sum_{a=1}^{2}\,m_a\,\int\limits_{C_a}^{}\!\!ds_a\,{,}
\end{equation}
where the new parameters $r_s$ and $\alpha$ are, respectively,
 the radius of gluonic tube and a dimensionless constant;
 $\Delta$ is the Laplace-
Beltrami operator for the induced metric $g_{ij}$
\begin{equation}
\Delta\,=\,\frac{1}{\sqrt{-g}}\,\frac{\partial}{\partial\xi^i}\,
\left(\sqrt{-g}\,g^{ij}\,\frac{\partial}
{\partial\xi^{j}}\right)\,{.}
\end{equation}
In the Gauss parametrization (2.2), the operator (5.2), up to the
second order in ${\bf u}$, can be written as
\begin{equation}
\Delta\,\simeq\,\Box\,+\,O({\bf u}^{2})\,{,}
\end{equation}
Now action (5.1) reads
$$
S\,\simeq \,-M_0^2\,\int\limits_{t_1}^{t_2}
\!\!dt\int\limits_{0}^{R}\!\!dr\,
\left[1-\frac{1}{2}\,\dot{\bf u}^2\,+\,\frac{1}{2}\,{\bf u}^{'2}\,
+\,\frac{\alpha}{2}\,
r_s^2\,(\Box{\bf u})^2\right]\,-$$
\begin{equation}
\,-\,\sum_{a=1}^{2}\,\frac{m_a}{2}\,\int
\limits_{t_1}^{t_2}\!\!dt\,\dot{\bf u}^2(t,r_a),
\quad r_1\,=0,\quad r_2\,=\,R\,{.}
\end{equation}

The equations of motion and boundary conditions for the
action (5.4) are
\begin{equation}
(1\,+\,\alpha\,r_s^2\,\Box)\,\Box\,{\bf u}\,=\,0\,{,}
\end{equation}
\begin{equation}
(1\,+\,\alpha\,r_s^2\,\Box)\,{\bf u}'\,=\,\frac{m}{M_0^2}\,
\ddot{\bf u}\,{,}
\quad r\,=0\,{,}
\end{equation}
\begin{equation}
(1\,+\,\alpha\,r_s^2\,\Box)\,{\bf u}'\,=\,-\,\frac{m}{M_0^2}\,
\ddot{\bf u}\,{,}
\quad r=R\,{,}
\end{equation}
\begin{equation}
\Box\,{\bf u}\,=\,0\,{,}\quad r\,=\,0,R,
\end{equation}
($m_1\,=\,m_2\,=\,m$). The Lagrangian in the action (5.4) depends
on the first and the second derivatives of the string coordinates,
therefore the number of obtained boundary conditions is twice
compared with the Nambu-Goto string.

The boundary value problem (5.5)--(5.8) reduces to the two
independent boundary problems. Indeed equations of motion are
given by product of commuting differential operators
$(1\,+\,\alpha\,r_s^2\,\Box)$ and $\Box$. Hence, the general
solution to this equations can be represented as a sum of two
terms
\begin{equation}
{\bf u}(t,r)\,=\,{\bf u}_1(t,r)\,+\,{\bf u}_2(t,r)\,{,}
\end{equation}
where
\begin{equation}
\Box\,{\bf u}_1\,=\,0\,{,}
\end{equation}
\begin{equation}
{\bf u}_1 '\,=\,-\,\frac{m}{M_0^2}\,\ddot{\bf u}_{1}
\,{,}\quad r\,=\,0\,{,}
\end{equation}
\begin{equation}
{\bf u}_1 '\,=\,+\,\frac{m}{M_0^2}\,\ddot{\bf u}_{1}
\,{,}\quad r\,=\,R\,{,}
\end{equation}
and
\begin{equation}
(1\,+\,\alpha\,r_s^2\,\Box)\,{\bf u}_2\,=\,0\,{,}
\end{equation}
\begin{equation}
{\bf u}_2(t,0)\,=\,{\bf u}_2(t,R)\,=\,0\,{.}
\end{equation}

In this case, ${\bf u}_1(t,r)$ is the solution for the Nambu-Goto
string with massive ends that we have analyzed in Section 2. The
string rigidity is taken into account by function ${\bf
u}_2(t,r)$. The general solution to Eq.~(5.13) obeying (5.14) can
be presented as
\begin{equation}
u_2^j(t,r)\,=\,i\,\frac{\sqrt{2}}{M_0}\,
\sum_{n\not= 0}^{}\,\exp\left[i\frac{
\nu_n}{R}\,t\right]\,\frac{\beta_n^j}{\nu_n}\,v_n(r)\,{,}\qquad
j\,=\,1,2,\ldots,D-2\,{.}
\end{equation}
The eigenfunctions $v_n(r)$ are given by
\begin{equation}
v_n(r)\,=\,-\,v_{-n}\,=\,N_n^{'}\,\sin\,n\pi\,\frac{r}{R}\,
{,}\quad n\,=\,1,2,\ldots
\,{,}
\end{equation}
where $N_n^{'}$ are normalization constants. For the natural frequencies,
$\nu_n$, in (5.15) we have
\begin{equation}
\nu_n\,=\,-\,\nu_{-n}\,=\,\sqrt{(\pi n)^2\,+\,
\frac{R^2}{\alpha r_s^2}}\,{,}
\qquad  n\,=\,1,2,\ldots\,{.}
\end{equation}
The amplitudes $\beta_n^j$ satisfy the usual relations of complex
conjugation $\beta_n^*\,=\,\beta_{-n},\, n=1,2,\ldots\,{.}$

The Hamiltonian formulation of the model under consideration is
developed in the following way. According to Ostrogradskii
{}~\cite{OSTR,ENKO} the canonical variables are defined by
\begin{equation}
q_1^j\,=\,u^j\,{,}\qquad q_2^j\,=\,\dot u^j\,{,}
\end{equation}
\begin{equation}
p_1^j\,=\,\frac{\partial L_{tot}}{\partial\dot u^j}\,
-\,p_2^j\,{,}\quad
p_2^j\,=\,\frac{\partial L_{tot}}{\partial\ddot u^j}\,{,}
\quad j\,=\,1,2,\ldots,D-2\,{,}
\end{equation}
where $L_{tot}$ is the Lagrangian density in action (5.4)
\begin{equation}
L_{tot}\,=\,\frac{M_0^2}{2}\,\left[\varepsilon(r)\,\dot{\bf u}^2\,
-\,{\bf u}^{'2}
\,-\,\alpha\,r_s^2\,(\Box{\bf u})^2\right]\,{.}
\end{equation}
Putting $L_{tot}$ and (5.9) into (5.18), (5.19) and taking into account
Eqs.~(5.10) and
(5.13) one obtains
\begin{equation}
\bf{q}_1 \,=\,{\bf u}_1 \,+\,{\bf u}_2\,{,}\quad \bf{q}_2\,
=\,\dot{\bf u}_1\,
+\,\dot{\bf u}_2\,{,}
\end{equation}
\begin{equation}
{\bf p}_1\,=\,M_0^2\,[\varepsilon(r)\,+\,\alpha\,r_s^2\,
\Box]\,\dot{\bf u}\,{,}
\quad {\bf p}_2\,=\,-\,\alpha\,r_s^2\,M_0^2\,\Box\,{\bf u}\,=\,M_0^2\,
{\bf u}_2\,{.}
\end{equation}

The canonical Hamiltonian is defined by
\begin{equation}
H\,=\,\int\limits_{0}^{R}\!\!dr\,[{\bf p}_1\dot{\bf q}_1\,+
\,{\bf p}_2\dot{\bf q}_2\,-\,L_{tot}]\,{,}
\end{equation}
In terms of Fourier amplitudes it becomes
\begin{equation}
H\,=\,\frac{1}{2\,R}\,\sum_{n=1}^{\infty}\,
\sum_{j=1}^{D-2}\,(\alpha_{nj}
\alpha_{nj}^{+}\,+\,\alpha_{nj}^{+}\alpha_{nj})\,-\,\frac{1}{2R}\,
\sum_{n=1}^{\infty}\,\sum_{j=1}^{D-2}\,(\beta_{nj}\beta_{nj}^{+}\,+\,
\beta_{nj}^{+}\beta_{nj})\,{.}
\end{equation}

The quantum theory is based on the canonical commutation relations
\begin{equation}
[u_a^i(t,r),\,p_b^j(t,r')]\,=\,i\,\delta_{ab}\,
\delta^{ij}\,\,\delta(r-r')\,{,}
\quad a\,=\,1,2\,{,}\quad i,j\,=\,1,2,\ldots,D-2\,{,}
\end{equation}
or in terms of the Fourier amplitudes
\begin{equation}
[\alpha_n^i,\alpha_m^j]\,=\,\omega_n\,\delta^{ij}
\,\delta_{n+m,0}\,{,}
\quad [\beta_n^i,\beta_m^j]\,=\,\nu_n\,\delta^{ij}\,\delta_{n+m,0}
\,{,}
\quad n,m\,=\,\pm 1,\pm2, \ldots\,{.}
\end{equation}

By introducing in standard way the annihilation and
creation operators
$$ a_n^i\,=\,(\omega_n)^{-1/2}\,\alpha_n^i\,{,}\qquad a_n^{i+}\,=\,
(\omega_n)^{-1/2}\,\alpha_n^{i+}\,{,} $$
\begin{equation}
b_n^j\,=\,(\nu_n)^{-1/2}\,\beta_n^j\,{,}\qquad b_n^{j+}\,
=\,(\nu_n)^{-1/2}
\,\beta_n^{j+}\,{,}
\end{equation}
$$ n\,=\, 1,2,\ldots\,{,}\qquad i,j,\,=\,1,2,\ldots,D-2\,{,}$$
the Hamiltonian operator (5.24) acquires the following form
$$
H\,=\,\frac{1}{R}\,\sum_{n=1}^{\infty}\,\omega_n\,
\sum_{j=1}^{D-2}\,a_n^{j+}
\,a_n^j\,-\,\frac{1}{R}\,\sum_{n=1}^{\infty}\,\nu_n\,
\sum_{j=1}^{D-2}\,
b_n^{j+}\,b_n^j\,+$$
\begin{equation}
\,+\,\frac{D-2}{2R}\,\left(\sum_{n=1}^{\infty}\,\omega_n\,
-\,\sum_{n=1}^{\infty}\,\nu_n\right)\,{.}
\end{equation}
The last two terms in (5.28) define the  Casimir  energy  in  the
model  under  consideration~\cite{NES}.  It  is important to note
that  the  second  oscillation  mode  with  frequencies  $\nu_n$,
responsible   for   the   string   rigidity,   gives  a  negative
contribution to the energy as compared with  the  oscillation  of
the first mode with frequencies $\omega_n$. This is also true for
the Casimir energy (see the last two terms in (5.28)).  It  is  a
direct  consequence  of  the  classical  expression for the total
energy in the rigid string model (5.24).  We point out that  this
defect is typical in all the theories with higher derivatives. To
remove it,  certainly in formal way only, the quantum states with
negative norm are used sometimes~\cite{EW,PEU}.

Now we   calculate   the   rigid  string  potential  in  one-loop
approximation.  Again we shall treat the Euclidean version of the
model under consideration.

The interquark potential is given by Eq. (3.2) with the Euclidean
action
$(t_1\,=0\,,\,t_2\,=T)$
\begin{equation}
S_E\,=\,M_0^2\,\int\limits_{0}^{T}\!\!dt\,\int\limits_{0}^{R}\!\!dr\,
\left[
1\,+\,\frac{1}{2}\,{\bf u}\,(1\,+\,\alpha\,r_s^2\,\partial^2)\,
\partial^2\,{\bf u}\,
\right]\,{.}
\end{equation}
As in section 4, the boundary terms in action (5.4) will be taken
into  account  by  finding  the   proper   eigenvalues   of   the
corresponding differential operator.

After functional integration, the potential takes the form
\begin{equation}
V(R)\,=\,\lim_{T\rightarrow\infty}\,\frac{1}{T}\,
\left[\int\limits_{0}^{T}\!\!
dt\int\limits_{0}^{R}\!\!\,dr\,M_0^2\,+\,\frac{D-2}{2}\,\mbox{Tr}
\ln G^{-1}\right]\,{,}
\end{equation}
where the      operator
$G^{-1}\,=\,(1\,+\,\alpha\,r_s^2\,\partial^2)\,\partial^2$
corresponds to the inverse of the Euclidean Green function
\begin{equation}
(1\,+\,\alpha\,r_s^2\,\partial^2)\,\partial^2\,G(x,x')\,=\,
\delta(x,x')\,{.}
\end{equation}

In view of (3.7), the functional trace in Eq. (5.30) can be
written as
$$
\mbox{Tr}\ln G^{-1}\,=\,\mbox{Tr}\ln\partial^2\,+\,\mbox{Tr}
\ln (1\,+\,\alpha\,r_s^2\,\partial^2)\,=
$$
\begin{equation}
=\,T\,\sum_{n}^{}\,\int\limits_{-\infty}^{+\infty}\!\!
\frac{d\omega}{2\pi}\,
\ln(\omega^2\,+\,k_n^{(1)\,2})\,+\,
T\,\sum_{n}^{}\,\int\limits_{-\infty}^{+\infty}\!\!
\frac{d\omega}{2\pi}\,
\ln\,[1\,+\,\alpha\,r_s^2\,(\omega^2\,+\,k_n^{(2)\,2})]\,{.}
\end{equation}
$k_n^{(1)}$ are the admissable values of the wave vector for  the
field  ${\bf  u}_1  (t,r)$  determined by the boundary conditions
(5.11)            and            (5.12),            $k_n^{(1)}\,=
\,\omega_n/R,\;n\,=\,1,2,\ldots$,   and   $k_n^{(2)}$   are   the
admissable  values  of  the  wave  vector  for  the  field  ${\bf
u}_2(t,r)$    determined    by    boundary   conditions   (5.14),
$k_n^{(2)}\,=\,\nu_n/R,\;n\,=\,1,2,\ldots$.    By    using    the
analytical  regularization,  formula  (3.9),  Eq.~(5.32)  can  be
reduced to
\begin{equation}
\mbox{Tr}\ln G^{-1}\,=\,\frac{2T}{R}\,v(q)\,+\,\frac{T}{R}\,\pi\,
\sum_{n=1}^{\infty}\,\sqrt{n^2\,+\,\epsilon^{-1}}\,{,}
\end{equation}
where the dimensionless parameter $\epsilon$ is defined by
\begin{equation}
\epsilon\,=\,\frac{\alpha}{2}\,\pi^2\,
\left(\frac{r_s}{R}\right)^2\,{,}
\end{equation}
and the function $v(q)$ is given in (3.11). Now the interquark
potential (5.30) assumes the form
\begin{equation}
V(R)\,=\,M_0^2\,R\,+\,\frac{D-2}{R}\,\left[v(q)\,+\,\frac{\pi}{2}\,
\sum_{n=1}^{\infty}\,\sqrt{n^2\,+\epsilon^{-1}}\right]\,{.}
\end{equation}
In Eq.~(5.35) the second mode of  oscillation  gives  a  positive
contribution to the energy. As was mentioned above, it means that
the formalism,  applied here,  effectively uses for excitation of
the second mode quantum states with negative norm.

To remove the divergences in Eq.~(5.35) one has to use a subtraction
procedure. The function $v(q)$  has been renormalized in Section~3
(Eqs.~(3.22) and (3.26)). Now we renormalize only the last sum in (5.35).
It can be rewritten in the following way
\begin{equation}
w(\epsilon)\,=\,\frac{1}{2}\,\sum_{n=1}^{\infty}\,\sqrt{n^2\,+\,
\epsilon^{-1}}\,
=\,\frac{1}{4}\,\left(\sum_{n=-\infty}^{+\infty}\,\sqrt{n^2\,+\,
\epsilon^{-1}}
\,-\,\frac{1}{\sqrt{\epsilon}}\right)\,{,}
\end{equation}
where the last term removes the term with $n\,=\,0$ in  the  sum.
The  renormalized  function  $w^{ren}(\epsilon)$ is obtained from
equation  (5.36)  by   subtracting   its   value   for   infinite
string~\cite{KCN}
$$w^{ren}(\epsilon)\,=\,w(\epsilon)\,-\,
\lim_{R\rightarrow\infty}\,w(\epsilon)\,=$$
\begin{equation}
\,=\frac{1}{4}\,\left(\sum_{n=-\infty}^{+\infty}\,-\,\int
\limits_{-\infty}^{+\infty}\!\!dn\right)\,\sqrt{n^2\,+\,
\epsilon^{-1}}\,-\,
\frac{1}{4\sqrt{\epsilon}}\,{.}
\end{equation}

Equation (5.37) can be expressed in terms of the
function~\cite{NES}
\begin{equation}
S(x)\,=\,\frac{1}{2}\,\left(\sum_{n=-\infty}^{+\infty}\,-\,\int
\limits_{-\infty}^{+\infty}\!\!dn\right)\,\sqrt{n^2\,+x}\,=\,
-\frac{\sqrt{x}}
{\pi}\,\sum_{n=1}^{\infty}\,\frac{1}{n}\,K_1(2\pi n\sqrt{x})\,{,}
\end{equation}
where $K_1(z)$ is the modified Bessel function ~\cite{GASR}. This
representation for $S(x)$ is  convenient  for  investigating  its
behaviour  at  large  $x$,  because  the modified Bessel function
decays  exponentially  ~\cite{GASR}.  For   small   $x$   another
representation of the function $S(x)$ was proposed~\cite{KCN}. By
using the definition of the function $S(x)$, Eq.~(5.37) becomes
\begin{equation}
w^{ren}(\epsilon)\,=\,\frac{1}{2}\,S(\epsilon)\,-\,
\frac{1}{4\sqrt{\epsilon}}\,{.}
\end{equation}

Finally one-loop calculation of the interquark potential gives
\begin{equation}
V(R)\,=\,M_0^2\,R\,+\,\frac{D-2}{R}\,\left[v^{ren}(q)\,-\,\frac{\pi}{4
\sqrt{\epsilon}}\,-\,\frac{1}{2\sqrt{\epsilon}}\,\sum_{n=1}^{\infty}\,
\frac{1}{n}\,K_1\left(\frac{2\pi n}{\sqrt{\epsilon}}\right)\right]\,{,}
\end{equation}
which depends on the quark mass and string rigidity  through  the
dimensionless parameters $q$ and $\epsilon$, respectively.

In the asymptotic limit, $R\rightarrow\infty$, the string potential (5.40)
assumes the following form
$$V(R) \simeq M_0^2\,R\,+\,
(D-2)\,\left(\frac{\ln 2}{\pi}\frac{M_0^2}{m}\,-\,\frac{\pi(D-2)}{24R}\,
-\,\frac{1}{2\sqrt{2\alpha}\,r_s}\right)\,+ $$
\begin{equation}
+\,\frac{D-2}{2\pi}\frac{m}{M_0^2}\,\frac{1}{R^2}\,-\,(D-2)\,\sqrt{\frac{2}
{\pi\alpha r_s}}\,\frac{1}{\sqrt{R}}\,
\exp[-2\sqrt{2}\,R/\sqrt{\alpha}r_s]\,{,}
\quad R\to\infty\,{.}
\end{equation}

The first  term  in  the  right--hand--side of (5.41) is linearly
rising potential,  $M_0^2 R$;  the  second  term  is  a  constant
determined by quark mass $m$, string tension $M_0^2$ and coupling
constant $\sqrt{\alpha}\,r_s$ in the rigid  string  model  action
(5.1);  the  third term is the universal L\"{u}scher $1/R$--term,
the fourth term is the mass  quark  correction,  $(D\,-\,2)m/2\pi
M_0^2\,R^2$.  The last term in (5.41) vanishes exponentially when
$R\to\infty$.   For   different   values   of   the    parameters
$M_0,m,\alpha$  and  $r_s$ the total constant term in (5.41) may,
in principle, change sign.

\section{Conclusion}

In this   paper   we  have  developed  a  consistent  method  for
calculating the interquark potential  generated  by  relativistic
string with point--like masses (spinless quarks) at its ends. The
obtained results indicate that the correction  to  the  potential
due  to the finite quark masses turns out to be considerable near
the critical (or deconfinement) distance. When the quark mass $m$
decreases the contribution of this correction rises. However, the
formula obtained cannot be obviously used at very small $m$.  The
point  is  that  all  our corrections are based on the linearized
dynamical equations in the  string  theory  which  are,  in  some
sense,  equivalent  to  the  nonrelativistic approximation in the
initial string action~\cite{VVN,NES}.  At small $m$,  when  large
velocities  of  the string ends are important\footnote{It is well
known~\cite{BARB} that free (massless) ends  of  the  Nambu--Goto
string    are   permanently   moving   with   light   velocity.},
nonrelativistic approximation is certainly not applicable.

The extension of the proposed method for investigating the  model
of   the   relativistic   string  with  massive  ends  at  finite
temperature is of undoubted interest.

\section{Acknowledgments}

V.V.N. thanks  Prof.~H.~Kleinert  and   Dr.~A.M.~Chervyakov   for
valuable discussion of some topics considered in this paper. G.L.
would like to thank Prof.~V.V.  Nesterenko and the members of the
Bogoliubov  Laboratory  of  Theoretical  Physics (JINR) for their
kind hospitality during his staying in Dubna. He also acknowledge
the financial support from JINR and the Salerno University.  This
work has been partly supported  by  the  Russian  Foundation  for
Fundamental Research through Project No. 93--02--3972.

\appendixa
          Here we calculate the sum (3.20) by making use of the
counter integral (3.24). Putting in (3.24) $f(z)\,=\,\sin\omega$
and deforming counter $C$ as was explained in Section 3, we
obtain

\begin{equation}
J\,=\,\frac{\pi}{2}\,\sum_{n=1}^{\infty}\,n\,=\,-\frac{1}{2\pi}\,
\int\limits_{0}^{+\infty}\!\!dy\,y\,
\coth y\,{.}
\end{equation}

The integral (A.1) diverges at the upper limit. To regularize it
we introduce into the integrand the cutting multiplier
$e^{-\epsilon y}$, $\epsilon>0$. After doing the integral we take
the limit $\epsilon\to 0$
\begin{equation}
J^{ren}\,=\,-\frac{1}{2\pi}\,\lim_{\epsilon\to 0}\,J_{\epsilon}\,{,}
\end{equation}
where
\begin{equation}
J_{\epsilon}\,=\,\int\limits_{0}^{\infty}\!\!dy\,y\,e^{-\epsilon y}
\coth y\,=\,\Gamma(2)\,\left[\frac{1}{2}\,\zeta\left(2,\frac{\epsilon}{2}
\right)\,-\,\frac{1}{\epsilon^2}\right]\,{,}
\end{equation}
and $\zeta(z,q)$ is the Hurwitz zeta-function ~\cite{GASR}
\begin{equation}
\zeta(z,q)\,=\,\sum_{n=0}^{\infty}\,\frac{1}{(n+q)^z}\,{.}
\end{equation}

Substituting (A.3) into (A.2) we obtain
\begin{equation}
J^{ren}\,=\,-\frac{\pi}{24}\,+\,\lim_{\epsilon\to 0}\,\left[-\frac{1}{2\pi
\epsilon^2}\right]\,{.}
\end{equation}
The pole in (A.5) should be dropped as it is usually done in analytical
regularization. Hence
\begin{equation}
J^{ren}\,=\,-\,\frac{\pi}{24}\,{.}
\end{equation}

Application of (3.24) to the fraction in (3.25) alone results,
 after deforming the counter $C$ and integrating by parts, in
 (3.27) without any divergences.

\appendixb
In this appendix the asymptotic expansion for $q\to\infty$
of the function $I(q)$ entered in $v^{ren}(q)$ will be obtained.
In terms of the variable $x\,=\,q^{-1}\,y$, Eq.~(3.27) reads
\begin{equation}
I(q)\,=\,\frac{q}{2\pi}\,\int\limits_{0}^{\infty}
\!\!dx\,\ln\left(1\,+\,\frac{1}{x^2\,+\,2x\,\coth qx}\right)\,{.}
\end{equation}
Let us divide the range of integration into two regions, $(0,1/q]\,
\cup\,[1/q,
\infty)$, with $q\to\infty$. The integral in (B.1) can be written as
$$
I(q)\,=\,\frac{q}{2\pi}\int\limits_{0}^{1/q}
\!dx\,\ln\left(1\,+\,\frac{1}{x^2\,+\,2x\,\coth qx}\right)\,+$$
$$
+\,\frac{q}{2\pi}
\,\int\limits_{1/q}^{\infty}\!dx\,\ln\left(1\,+\,\frac{1}{x^2\,+\,2x\,
\coth qx}\right)\,=
$$
\begin{equation}
=\,\frac{q}{2\pi}\,J_1(q)\,+\,\frac{q}{2\pi}\,J_2(q)\,{.}
\end{equation}
For $x\in(0,1/q]$ we can substitute $\coth qx$ by $(qx)^{-1}$ and the
integral $J_1(q)$ in (B.2) gives
\begin{equation}
J_1(q)\simeq \int\limits_{0}^{1/q}\!dx\,\ln\left(1\,+\,\frac{1}
{x^2\,+\,2/q}\right)\,\simeq\,\frac{1}{q}\,\ln\frac{q}{2}\,+\,
\frac{1}{q^2}\,{.}
\end{equation}
When $x\in [1/q,\infty)$ the function $\coth qx$ can be
approximate by 1, so that the integral $J_2(q)$ in (B.2) assumes the
following form
\begin{equation}
J_2(q)\simeq \int\limits_{1/q}^{\infty}\!dx\,\ln\left(1\,+\,\frac{1}
{x^2\,+\,2x}\right)\simeq 2\,\ln 2\,-\,\frac{1}{q}\,\ln\frac{q}{2}\,{.}
\end{equation}
Putting Eqs.~(B.3) and (B.4) into (B.2) yields
\begin{equation}
I(q)\simeq \,\frac{\ln 2}{\pi}\,q\,+\,\frac{1}{2\pi\,q}\,
{,}\quad q\to\infty
\,{.}
\end{equation}

\newpage
\vspace*{3cm}
\centerline{\large \bf Figure Captions}
\vskip2cm
Fig.~1. Renormalized sum over the natural frequencies  of  the
string with massive ends,  $v^{ren}(q)$,  calculated by making
use of  the  definition  (3.22)  and the  integral  representation
(3.26),   (3.27).   The  results  are,  certainly,  the  same.
\vskip0.8cm

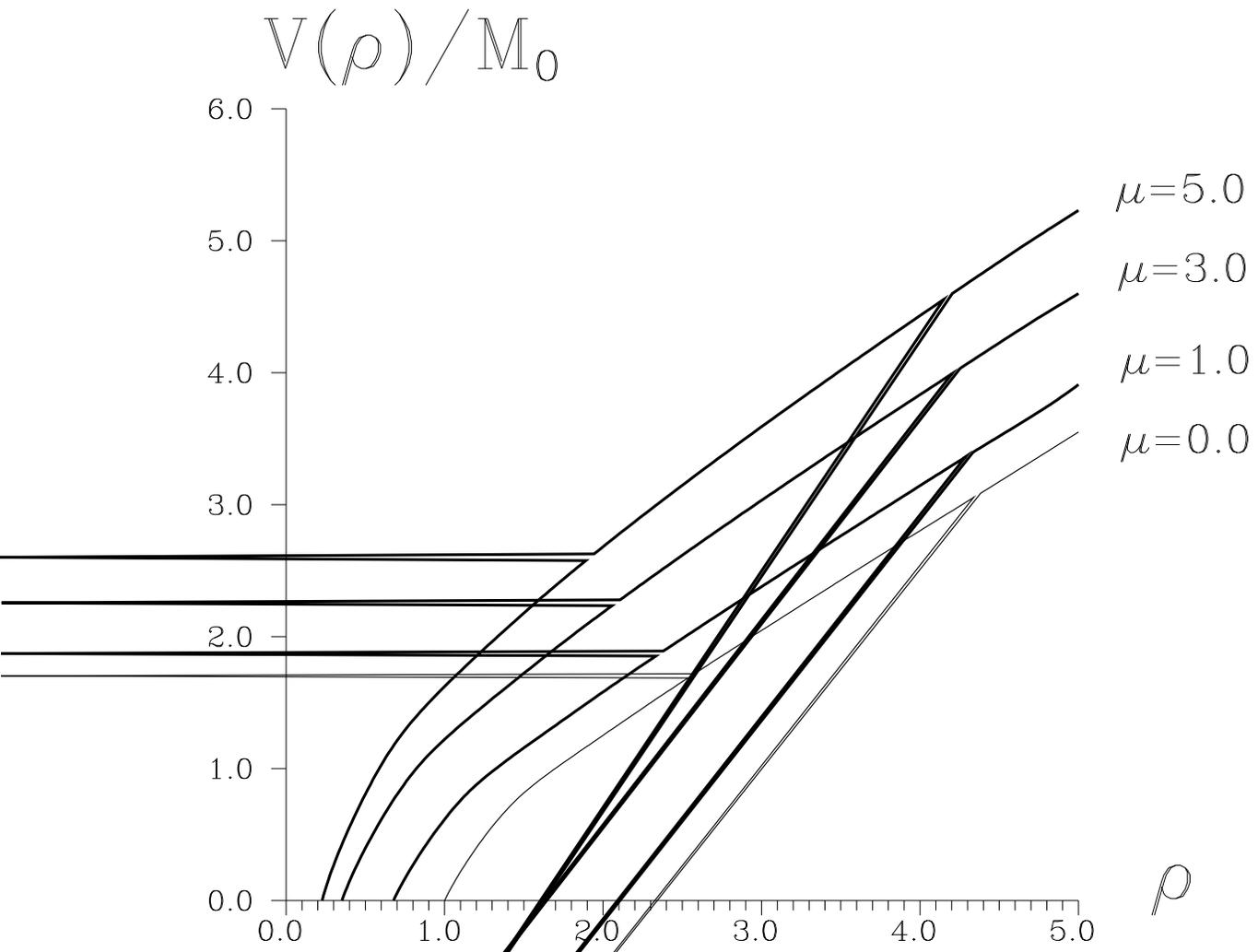
Fig.~2. Dimensionless    interquark    potential    $V(R)/M_0$
calculated through Eq.~(4.10) for different  values  of  ratio
$\mu\,=\,M_0/m$.  The  Alvarez  result~[1] for the Nambu--Goto
string with fixed ends is obtained at $\mu\,=\,0$.

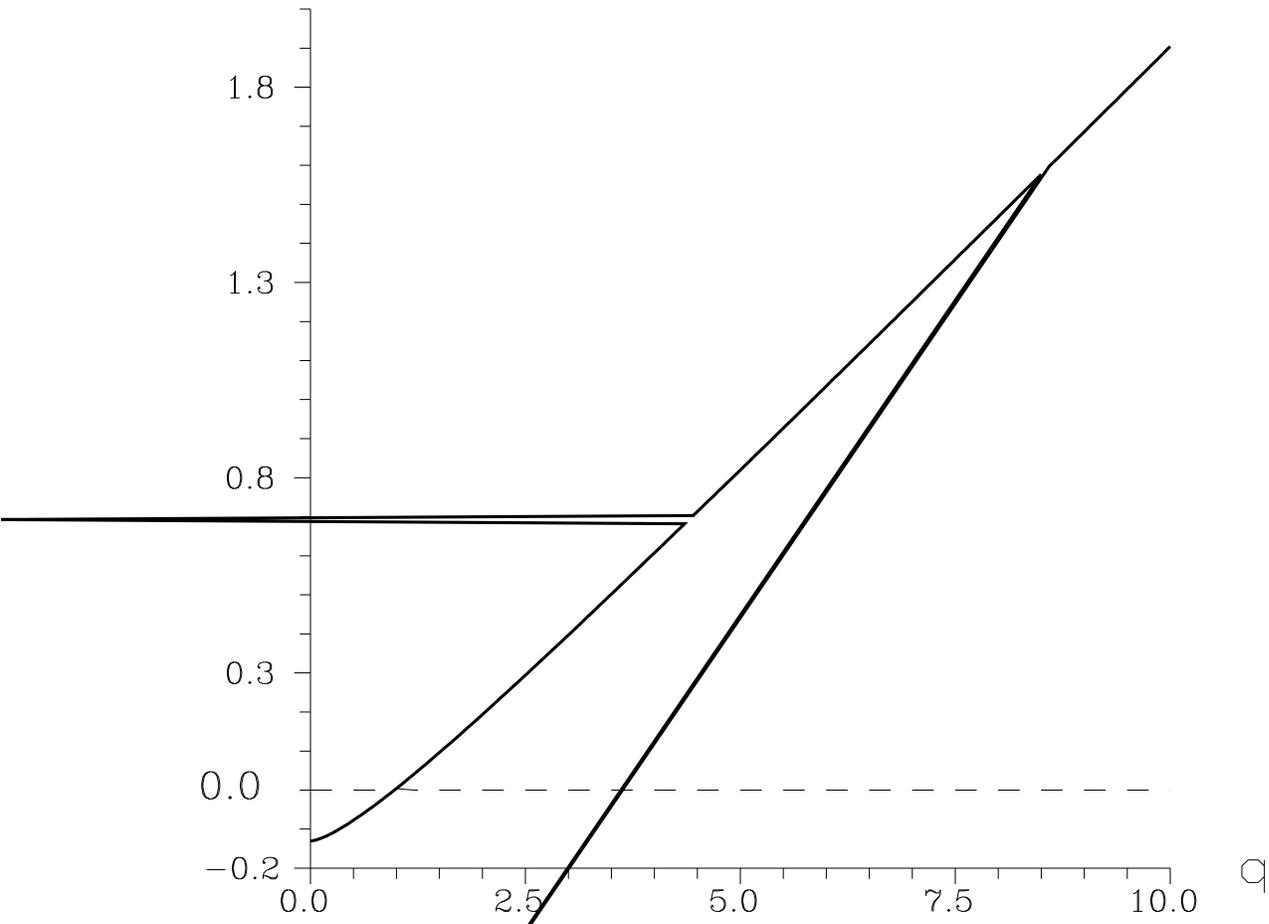
\end{document}